Quasi-free-standing monolayer hexagonal boron nitride on Ni

Satoru Suzuki[1], Yuichi Haruyama[1], Masahito Niibe[1], Takashi Tokushima[1], Akinobu Yamaguchi[1], Yuichi Utsumi[1], Atsushi Ito[2], Ryo Kadowaki[3], Akane Maruta[3], and Tadashi Abukawa[3]

[1]Laboratory of Advanced Science and Technology for Industry, University of Hyogo, 3-1-2 Koto, Kamigori, Ako, Hyogo 678-1205, Japan
[2]School of Engineering & Graduate School of Engineering, University of Hyogo, 2167 Shosha, Himeji, Hyogo 671-2280, Japan
[3]Institute of Multidisciplinary Research for Advanced Materials, Tohoku University, 2-1-1 Katahira, Aoba, Sendai, Miyagi 980-8577, Japan

Abstract
The electronic structure of monolayer hexagonal boron nitride grown on Ni by the diffusion and precipitation method was studied by x-ray absorption spectroscopy, emission spectroscopy, x-ray photoelectron spectroscopy and micro-ultraviolet photoemission spectroscopy. No indication of hybridization between h-BN π and Ni 3d orbitals was observed. That is, the monolayer h-BN was found to be in the quasi-free-standing state. These results are in striking contrast to those of previous studies in which h-BN was strongly bound to the Ni surface by the orbital hybridization. The absence of hybridization is attributed to absence of a Ni(111) surface in this study. The lattice-matched Ni(111) surface is considered to be essential to orbital hybridization between h-BN and Ni.

1 Introduction
Hexagonal boron nitride (h-BN) has been attracting much attention as a substrate material for graphene devices [1-3], and as a barrier layer for carrier [4-6] and spin [7-9] tunneling devices. Now, monolayer h-BN is usually grown on transition metal substrates, such as Ni using chemical vapor deposition (CVD) [10-12]. In bulk h-BN, the neighboring layers are very weakly bound by the van der Waals interaction. However, previous studies using core level spectroscopies have shown that, in the CVD-grown h-BN/Ni system, there is strong hybridization between h-BN π and Ni 3d orbitals and that h-BN is chemisorbed on the Ni surface [13-17]. Similarly, indications of strong hybridization have been also observed in h-BN/Ni systems grown by molecular beam epitaxy (MBE)[18] and by ion beam sputtering deposition methods [19]. Theoretical calculations also support chemisorption of h-BN on Ni [20, 21]. Calculation even predicts that the large band gap of h-BN will disappear because of hybridization of the h-BN π state with Ni 3d at the Fermi level and that h-BN on Ni is metallic [21]. The nature of the bonding between h-BN and Ni would be particularly important to the heteroepitaxial growth of graphene on h-BN/Ni. Usachov et al. demonstrated freeing h-BN from a Ni substrate by intercalation of Au between h-BN and Ni [17]. They also succeeded in CVD growth of graphene onto the fully relaxed and physisorbed h-BN on Au/Ni. The existence or non-existence of hybridization would also be important for other physical and chemical properties, such as tunneling barrier, gas absorption, and surface electric properties.

The diffusion and precipitation method is another way to grow atomically thin h-BN films [22-25]. This method utilizes only a solid phase reaction. Thus, it is much safer than the CVD, which uses flammable and toxic gases as source and carrier gases. In this study, monolayer h-BN was grown on Ni by the diffusion and precipitation method, and the bonding character between the monolayer h-BN and Ni, which has not yet been addressed, was studied using x-ray absorption spectroscopy (XAS), x-ray emission spectroscopy (XES), x-ray photoelectron spectroscopy (XPS), and photoemission electron microscopy (PEEM). No indication of hybridization between h-BN π and Ni 3d orbitals was found in the results, in striking contrast to previous reports [13-19]. A possible reason for this is discussed.

2 Experiment

Submonolayer h-BN films were grown on Ni foil using the diffusion and precipitation method, based on our previous studies [23, 24] (See Fig. S1 in the Suplementary Data). Briefly, a ~100-nm-thick amorphous boron nitride (a-BN) film was deposited on a Ni foil (Nilaco Co., thickness: 25 μm) using the radio frequency magnetron sputtering method. The a-BN/Ni samples were heated either in the endstation of the beamline BL7B in the NewSUBARU synchrotron radiation facility or in a stand-alone PEEM apparatus. The heating temperature and time were about 900 °C and 30 minutes, respectively. During the heating, a fraction of B and N atoms of the a-BN film diffused through the Ni foil and formed h-BN on the opposite face of the Ni foil. The samples had been kept in an ultra-high vacuum (~$10^{-7}$ Pa or below) until XAS, XPS and PEEM measurements were completed, without otherwise mentioning.

Polarized B-$K$ and N-$K$ XAS measurements were performed at the beamline BL7B at NewSUBARU equipped with a constant deviation angle varied-line-spacing plane grating monochromator [26]. The resolving power $E/\Delta E$ was set to ~1000. The absorption intensity was measured by the partial electron yield method. XPS measurements were also performed in the same analysis chamber using an Al $K\alpha$ source and a hemispherical electron analyzer (Scienta, R-3000).

B-*K* and N-*K* XES measurements were performed ex-situ at the beamline BL9A at NewSUBARU equipped with a constant deviation angle varied-line-spacing plane grating monochromator [27] and a soft X-ray emission spectrometer [28]. The optical axis of the spectrometer was set perpendicular to the beamline axis in the plane of polarization of the incident light. The excitation photon energies were 210 and 420 eV for B-*K* and N-*K* XES, respectively. These energies are sufficiently higher than the absorption threshold, to allow the spectra to be obtained in the non-resonant mode. The resolving powers of the spectrometer were about 700 for measurements of monolayer h-BN/Ni and 1000 for bulk h-BN. The takeoff angle of emission was varied by rotating the sample. To our knowledge, this is the first observation of takeoff angle dependence of XES from monolayer h-BN. Accumulation times for the monolayer h-BN/Ni sample were 30 m for both B-*K* and N-*K* spectra. This value is much shorter than that of a previous XES measurement (10 h) reported on monolayer h-BN [16].

In-situ PEEM observation was performed in a PEEM apparatus (IS-PEEM, FOCUS GmbH) equipped with a micro-analyzer [29, 30]. Local work function measurements by selected-area ultraviolet photoemission spectroscopy (micro-UPS) were also performed using the same apparatus. A high-pressure mercury-vapor lamp (hν~4.9 eV) was used as the light source.

A field emission scanning electron microscope (SEM) (JEOL, JSM-7001F) was used for ex-situ electron backscatter diffraction (EBSD) measurements and SEM observation. The electron beam energy was 15 keV.

3 Results

3.1 B-*K* and N-*K* XAS

As shown in Fig. 1(a), bulk h-BN (unoriented powder sample) shows a sharp and strong exciton peak at about 192 eV in the B-*K* absorption spectrum, as denoted by A. This peak is caused by the transition from B 1s to π* ($2p_z$). The transition to σ* orbital, which contains $2p_x$ and $2p_y$ components, is observed at around 198 and 200 eV, as denoted by B and C. In the N-*K* spectrum, the transition from N 1s to π* orbital is observed at about 402 eV. The small hump is observed at about 404 eV as denoted by A''. Here, we use these denotations in accordance with ref. 13. The transitions to σ* orbitals are observed at around 407 and 409 eV, as shown in Fig. 1(b). The excitonic effect is much less prominent in the N-*K* spectrum, meaning that the unoccupied π* state has larger B $2p_z$ (smaller N $2p_z$) character. This can be basically explained in terms of the higher electronegativity of N than B; that is, a certain degree of ionic character of B-N bonds.

In previous XAS studies of h-BN/Ni systems, the B-*K* spectra have been found to be considerably different from those of bulk h-BN. At both lower and higher energies of the original exciton peak, additional broad spectral features appear (often denoted by A' and A'') [13-15, 17-19]. Alternatively, the sharp exciton peak A becomes broader and less intense. In the N-*K* spectrum, A' peak also appears at about 398 eV, at the low energy side of the threshold [13-15, 18, 19]. The A peak becomes less intense and the A'' peak becomes much more intense. Both the newly appeared features A' and A'' have been shown to have π* symmetry [13, 18]. Based on these results, it has been concluded that there is a large degree of hybridization between h-BN π (π*) and Ni 3d orbitals. Band structure calculations [20, 21] also support this result, as mentioned above.

Figure 1(a) also shows the polarized B-*K* spectrum of h-BN/Ni grown by the diffusion and precipitation method. A SEM image is of the sample is shown in Fig. 2. The dark and bright regions are considered to be h-BN and bare Ni surface, respectively, based on our previous combined analyses of micro-Auger and SEM [24]. From the figure, the coverage of monolayer h-BN was evaluated to be about 0.8. For a planar h-BN sheet, absorption intensities of π* and σ* symmetries are expected to be simply proportional to $\sin^2\theta$ and $\cos^2\theta$, respectively (θ: incidence angle). The large polarization dependence indicates that the h-BN atomic sheet is oriented parallel to the Ni surface, as reasonably expected. The incidence angle of 35° almost corresponds to the magic angle at which an unpolarized spectrum is obtained. In good contrast to previous studies, the spectrum of h-BN/Ni is very similar to that of bulk h-BN and the peaks A' and A'' do not appear. Similarly, the N-*K* spectra of h-BN/Ni is very similar to that of bulk h-BN and no additional features are observed. These results indicate that the monolayer h-BN grown by the diffusion and precipitation method on the Ni foil almost sustains the original electronic structure of h-BN.

## 3.2 B-*K* and N-*K* XES

B-*K* and N-*K* XES of bulk h-BN are shown in Fig. 3. Here, in the bulk h-BN sample, c axes of microcrystals are highly oriented by compressing [28]. In this case, considering both s- and p-polarized light emission, intensities of π and σ symmetries are expected to be proportional to $\cos^2\theta$ and $(1+\sin^2\theta)/2$, respectively [31, 32]. In contrast to XAS, a prominent π peak is observed at about 394 eV in N-*K* XES of bulk h-BN [28][33] and is not observed in B-*K* XES because the π electron more largely occupies the N $2p_z$ than the B $2p_z$ orbital because of the specific ionic character of B-N bonding. The other peaks in the N-*K* spectrum are attributed to σ orbitals [28, 33]. In the B-*K* spectrum, both π and σ spectra are broad and have considerable overlap. Roughly speaking, intensities at relatively high photon energies centered at 184 eV are largely contributed by π band and at lower energies around 182 eV by σ bands [32].

In a previous study on CVD-grown h-BN/Ni [16], a gap state due to h-BN π-Ni 3d hybridization was observed in the bandgap region (188-190 eV) in B-*K* XES. Changes observed in N-*K* XES were much more drastic. The π peak completely disappeared. Instead, the spectral intensity was mostly transferred to the gap state (396-398 eV).

XES of h-BN/Ni obtained at three incidence (take off) angles are also shown in Fig. 3. As shown in Fig. 3(a), B-*K* XES of h-BN/Ni largely differs from bulk h-BN and only a weak takeoff angle dependence is observed, strongly suggesting that a spectrum of some form of B other than h-BN is superposed (the sample was the same as the XAS and XPS measurements.). However, with the exception of a small amount of excess B and its oxide (Fig. 4), no such considerable amount of another form of B has been observed in surface sensitive techniques, such as XAS (Fig. 1), XPS (Fig. 4), micro-Auger spectroscopy [22-24], or in cross-sectional transmission electron microscope observation [22],. In fact, XES of metallic B [34] and $B_2O_3$ [35] do not explain the characteristic feature observed at 187.5 eV. We think that the additional spectrum comes from B atoms that had diffused in the Ni foil. As mentioned in our previous report, the solubility of B in Ni reaches 0.3 at.% at 1085°C [36], although the solubility of N is negligibly small even at high temperatures [37, 38]. The B atoms in Ni would remain inside when the sample was quenched after heating. Considering that XES is bulk-sensitive, it is not surprising that the number of B atoms in the probing depth is comparable to or larger than that of 0.8 ML of h-BN at the surface. In fact, XES has been used to study the electronic structure of dopant B atoms with a concentration of 0.03 % [34]. Now, we would like to focus on the bandgap region. No gap state is observed in the h-BN/Ni sample, suggesting the absence of h-BN π and Ni 3d hybridization.

N-*K* XES of the h-BN/Ni sample is similar to that of h-BN, as shown in Fig. 3(b). The slightly broader spectra are at least partly due to the differences in the resolving power (~700 for h-BN/Ni and 1000 for bulk h-BN) of the x-ray spectrometer. The incidence angle dependences of the π and σ peaks are, again, consistent with the planar h-BN sheet. As clearly seen in the figure, the π peak still exists and the gap state does not appear.

These XAS and XES results are consistent with each other and indicate that there is little hybridization between h-BN π and Ni 3d and that h-BN is physisorbed on the Ni surface. In other words, the monolayer h-BN is in a quasi-free-standing state on Ni, in striking contrast to the findings in previous studies [13-19] in which h-BN was found to be strongly bound to Ni by the hybridization.

3.3 B1$s$ and N1$s$ XPS

The bonding characteristic (physisorption or chemisorption) between h-BN and Ni affects spectra obtained by other techniques, such as XPS. It is well known that surface of bulk h-BN is chemically stable and very inert. However, a remarkably large spectral shift caused by air exposure has been observed in XPS of h-BN grown by the MBE method [18]. The air exposure caused spectral shifts to the low binding energy side by about 1.0 eV for both B 1s and N 1s levels [18]. This suggests that the adsorbates are not simply physisorbed on the h-BN surface and that there is a considerable charge transfer between h-BN and adsorbates. The hybridization between h-BN and Ni 3d reduces the inertness of the h-BN surface. On the contrary, such an air exposure-induced spectral shift was not observed for h-BN grown by the diffusion and precipitation method as shown in Figure 4. The B 1s and N 1s levels are observed at the binding energies of 190.0 and 397.6 eV, and a spectral shift is not observed before and after air exposure (in the B 1s spectra, a small amount of excess B is observed at 188 eV before air exposure and oxide of the excess B is observed at 191.5 eV after air exposure). This indicates that adsorbates possibly caused by the air exposure are physisorbed on h-BN and that there is little charge transfer between adsorbates and h-BN. Moreover, the binding energy difference between B 1s and N 1s levels was determined to be 207.7 eV, which is exactly the same as that of bulk h-BN [13]. A slightly larger value of 208.1 eV has been reported for CVD-grown h-BN/Ni [13], suggesting a greater ionic character of the B-N bonding. These XPS results show that the h-BN sustains the original very inert electronic structure and is consistent with the lack of hybridization between h-BN π and Ni 3d.

3.4 Work function measurements

The work function of CVD-grown h-BN on Ni(111) (work function: 5.3 eV [14][39]) was reported to be 3.6 eV [14][39]. That is, h-BN growth decreases the work function by 1.7 eV. The large work function decrease strongly suggests electron transfer from h-BN to Ni and dipole formation at the h-BN/Ni interface.

Figure 5(a) shows a PEEM image of h-BN/Ni. Bare Ni surface and h-BN domains are observed darkly and brightly, respectively (See also Figure S1 in the Supplementary

Data). The contrast is mainly considered to be due to work function change induced by h-BN formation. The enhanced photoemission from the h-BN-covered surfaces indicates that h-BN formation also decreases the work function in our case. However, the work function decrease is less prominent than in previous studies. Figure 5(b) shows secondary electron spectra obtained from several points. Curves A and B were obtained from bare Ni surfaces, and curves C, D, and E were from h-BN/Ni. From the cutoff energies of curves A and B, the work functions of the Ni surfaces are evaluated to be about 5.0 and 4.8 eV. The difference is considered to be caused by the difference in orientation of Ni grains due to the polycrystalline structure of the Ni foil. The work functions of h-BN/Ni are evaluated to be 4.6, 4.5, and 4.0 eV from curves C, D, and E. These values are considerably larger than that in a previous report (3.6 eV [14,39]). The work function decreases are within a range of 0.2 to 1.0 eV, which also considerably lower than those in previous studies (1.7 eV). The relatively large work functions and small work function decreases suggest the absence of a large dipole at the h-BN/Ni interface and thus, the absence of large charge transfer between h-BN and Ni. Similarly, a relatively large work function of 4.7 eV has been reported for the h-BN/Cu(111) system, in which the hybridization is much weaker than that on Ni(111) [14].

4. Discussion

As we have shown above, no indication of hybridization between h-BN $\pi$ and Ni 3d was observed in the h-BN/Ni samples grown by the diffusion and precipitation method, in striking contrast to previous studies [13-19]. Here, we discuss the possible reason for this. In all the previous studies, single-crystalline Ni was used, and monolayer h-BN films were grown on Ni(111) surfaces [13-19]. However, the Ni foil used in this study had a polycrystalline structure with a typical grain size of 100 μm, as shown in Fig. 6. The $z$-direction EBSD map reveals that there is no (111) face after the h-BN formation.

The Ni(111) surface has a six-fold symmetry. Its lattice constant is 0.249 nm, which is close to that of h-BN (0.250 nm) [21]. In the h-BN/Ni(111) system, N atoms are located on top of every outermost Ni atom, and B atoms are at either the fcc or hcp site (in any case, B atoms are at the top of the center of the triangle formed by three outermost Ni atoms.) [40]. The distance between the N and Ni atoms is 0.21 nm [41]. The translational symmetry and the relatively close N-Ni distance may be essentially important for the chemisorption properties of h-BN/Ni(111).

In this study, the symmetry and lattice constant of the Ni surface do not match those of h-BN at all. The lack of translational symmetry makes a theoretical calculation difficult. In fact, there have been very few reports on the electronic structure of h-BN on

non-(111) Ni surfaces. Nonetheless, we think that the lack of hybridization is due to the absence of the Ni(111) surface in this study. That is, the hybridization between h-BN π and Ni 3d is characteristic of the Ni(111) surface, which matches atomic structure of h-BN. We could not conceive of any explanation for the fact that the difference in the growth methods results in the difference in bonding character. It would be possible to grow quasi-free-standing h-BN on Ni by growth techniques other than the diffusion and precipitation method by using a non-(111) Ni surface.

The absence of hybridization would greatly affect the heteroepitaxial growth of graphene or another 2D material on it. Our results also suggest that the physical and chemical properties of h-BN on Ni, such as tunneling barrier, gas absorption, and surface electric properties, can be controlled by changing the crystal orientation of Ni.

## 5. Conclusion

Submonolayer h-BN was grown on a polycrystalline Ni foil by the diffusion and precipitation method and its electronic structure was studied by XAS, XES, XPS, and micro-UPS. In contrast to previous reports, no indication of hybridization between h-BN π and Ni 3d orbitals was observed. The quasi-free-standing state of h-BN is considered to be due to the absence of the Ni(111) surface in our sample. The lattice-matched Ni(111) surface seems to be essential for hybridization between h-BN π and Ni 3d orbitals. Heteroepitaxial growth of graphene on h-BN/Ni and the physical and chemical properties of h-BN/Ni may strongly depend on the crystal orientation of the Ni surface.

Acknowledgements
We thank Mr. Masaya Takeuchi for his technical support with the a-BN deposition. This work was supported by JSPS KAKENHI (JP16H03835, JP16H06361).

References

[1] Dean C R *et al* 2010 *Nat Nanotechnol*. **5** 722

[2] Petrone N, Dean C R, Meric I, van der Zande A M, Huang P Y, Wang L, Muller D, Shepard K L, and Hone J 2012 *Nano Lett*. **12** 2751

[3] Wang J, Ma F, Liang W, and Sun M 2017 *Materials Today Physics*. **2** 6

[4] Britnell L *et al* 2012 *Nano Lett*. **12** 1707

[5] Britnell L *et al* 2012 *Science*. **335** 947

[6] Yadav D, Tombet S B, Watanabe T, Arnold S, Ryzhii V, and Otsuji T 2016 *2D Materials*. **3**

[7] Piquemal-Banci M *et al* 2016 *Appl Phys Lett*. **108** 102404

[8] Piquemal-Banci M, Galceran R, Martin M-B, Godel F, Anane A, Petroff F, Dlubak B, and Seneor P 2017 *Journal of Physics D: Applied Physics*. **50**

[9] Karpan V M, Khomyakov P A, Giovannetti G, Starikov A A, and Kelly P J 2011 *Phys Rev B*. **84** 153406

[10] Cho H, Park S, Won D I, Kang S O, Pyo S S, Kim D I, Kim S M, Kim H C, and Kim M J 2015 *Sci Rep*. **5** 11985

[11] Ismach A *et al* 2017 *2D Materials*. **4** 025117

[12] Lee Y-H *et al* 2012 *RSC Adv*. **2** 111

[13] Preobrajenski A B, Vinogradov A S, and Mårtensson N 2004 *Phys Rev B*. **70** 165404

[14] Preobrajenski A B, Vinogradov A S, and Mårtensson N 2005 *Surf Sci*. **582** 21

[15] Preobrajenski A B, Vinogradov A S, Ng M L, Ćavar E, Westerström R, Mikkelsen A, Lundgren E, and Mårtensson N 2007 *Phys Rev B*. **75** 245412

[16] Preobrajenski A B, Krasnikov S A, Vinogradov A S, Ng M L, Käämbre T, Cafolla A A, and Mårtensson N 2008 *Phys Rev B*. **77** 085421

[17] Usachov D, Adamchuk V K, Haberer D, Grüneis A, Sachdev H, Preobrajenski A B, Laubschat C, and Vyalikh D V 2010 *Phys Rev B*. **82** 075415

[18] Tonkikh A A, Voloshina E N, Werner P, Blumtritt H, Senkovskiy B, Guntherodt G, Parkin S S, and Dedkov Y S 2016 *Sci Rep*. **6** 23547

[19] Meng J *et al* 2017 *Small*. **13** 1604179

[20] Miyashita A, Maekawa M, Wada K, Kawasuso A, Watanabe T, Entani S, and Sakai S 2018 *Phys Rev B*. **97** 195405

[21] Huda M N and Kleinman L 2006 *Phys Rev B*. **74** 075418

[22] Suzuki S, Pallares R M, and Hibino H 2012 *Journal of Physics D: Applied Physics*. **45** 385304

[23] Suzuki S, Molto Pallares R, Orofeo C M, and Hibino H 2013 *J Vac Sci Technol B*. **31** 041804

[24] Suzuki S, Ogawa Y, Wang S, and Kumakura K 2017 *Jpn J Appl Phys*. **56** 06GE06

[25] Chaohua Z, Lei F, Shuli Z, Yu Z, Hailin P, and Zhongfan L 2014 *Advanced Materials*. **26** 1776

[26] Kanda A, Sato T, Goto H, Tomori H, Takana S, Ootuka Y, and Tsukagoshi K 2010 *Physica C:*

*Superconductivity and its Applications*. **470** 1477

[27] Niibe M, Mukai M, Miyamoto S, Shoji Y, Hashimoto S, Ando A, Tanaka T, Miyai M, and Kitamura H 2004 *AIP Conf Proc*. **705** 576

[28] Niibe M and Tokushima T 2016 *AIP Conf Proc*. **1741** 030042

[29] Kadowaki R, Kuriyama M, Abukawa T, Sagisaka K, and Fujita D 2015 *e-Journal of Surface Science and Nanotechnology*. **13** 347

[30] Kadowaki R, Sano N, and Abukawa T 2017 *e-Journal of Surface Science and Nanotechnology*. **15** 115

[31] Mansour A, Schnatterly S E, and Carson R D 1985 *Phys Rev B*. **31** 6521

[32] Niibe M, Takehira N, and Tokushima T 2018 *e-Journal of Surface Science and Nanotechnology*. **16** 122

[33] Miyata N, Yanagihara M, Watanabe M, Harada Y, and Shin S 2002 *J Phys Soc Jpn*. **71** 1761

[34] Muramatsu H, Iihara J, Takabe T, and Denlinger J D 2008 *Analytical Sciences*. **24** 4

[35] Yumatov V D, Il'inchik E A, and Volkov V V 2003 *Russian Chemical Review*. **72** 1011

[36] Yang P C, Prater J T, Liu W, Glass J T, and Davis R F 2005 *Journal of Electronic Materials*. **34** 1558

[37] Kowanda C and Speidel M O 2003 *Scripta Materialia*. **48** 1073

[38] Abdulrahman R F and Hendry A 2001 *Metallurgical and Materials Transactions B*. **32** 1103

[39] Nagashima A, Tejima N, Gamou Y, Kawai T, and Oshima C 1995 *Phys Rev Lett*. **75** 3918

[40] Auwärter W, Muntwiler M, Osterwalder J, and Greber T 2003 *Surf Sci*. **545** L735

[41] Ohtomo M *et al* 2017 *Nanoscale*. **9** 2369

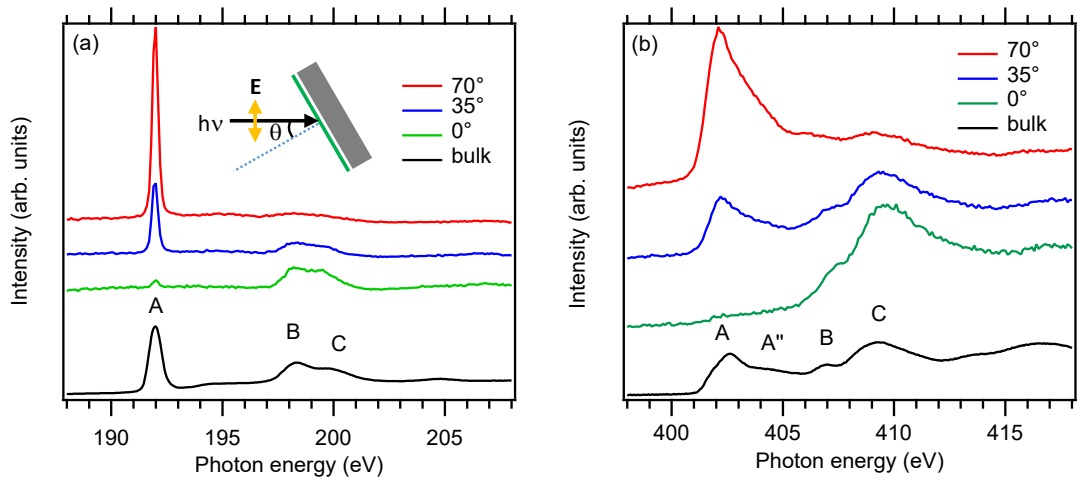

Fig. 1. Polarized (a) B-$K$ and (b) N-$K$ XAS of monolayer h-BN/Ni and bulk h-BN.

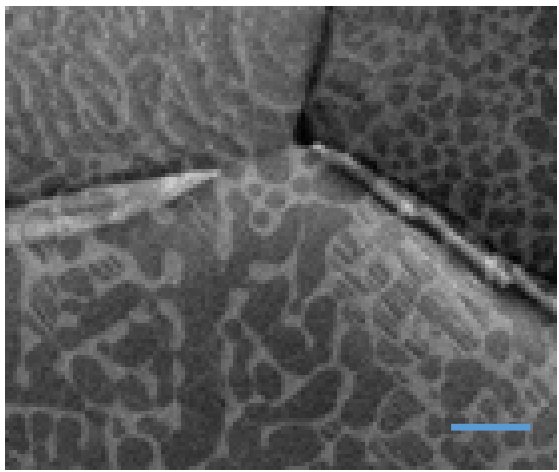

Fig. 2. SEM image of h-BN grown on Ni. Scale bar: 1 μm.

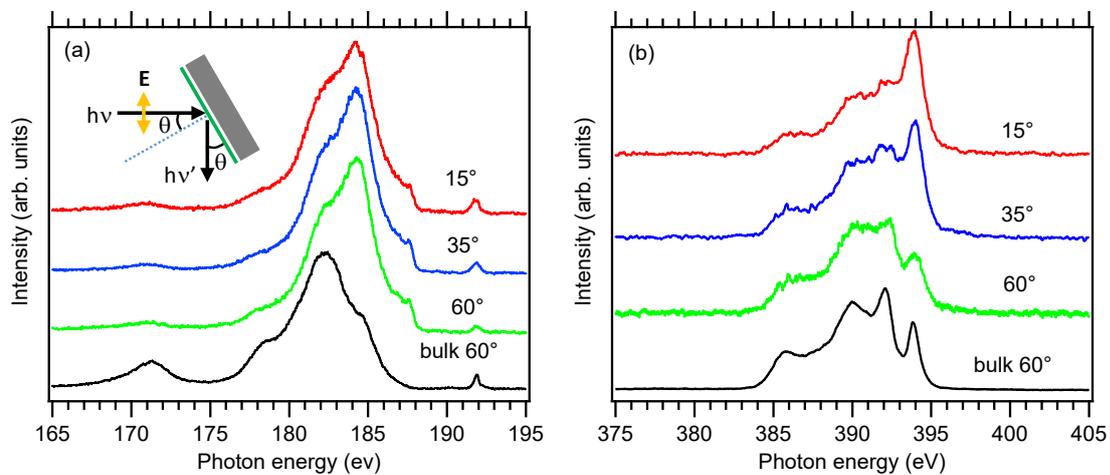

Fig. 3. (a) B-*K* and (b) N-*K* XES of monolayer h-BN/Ni and bulk h-BN.

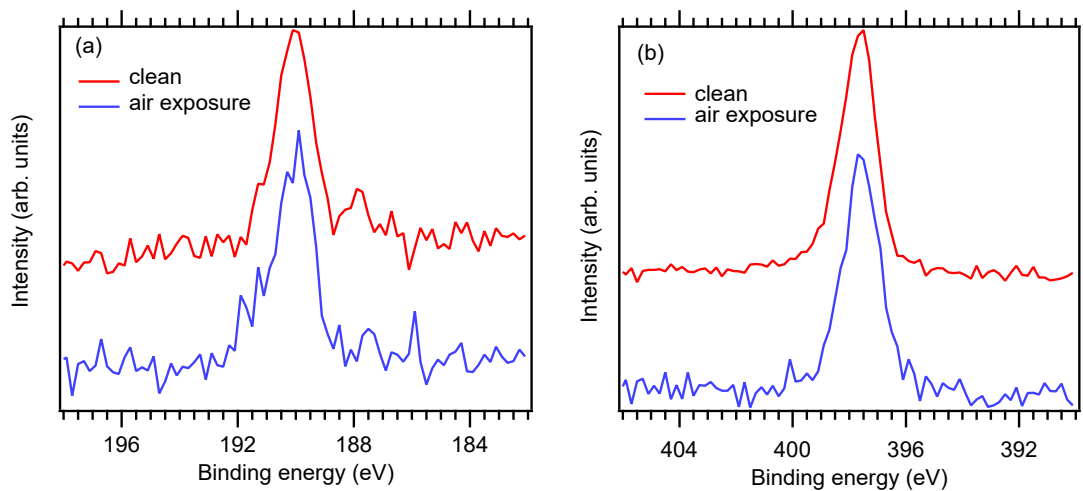

Fig. 4. (a) B 1s and (b) N 1s XPS of h-BN/Ni before and after air exposure.

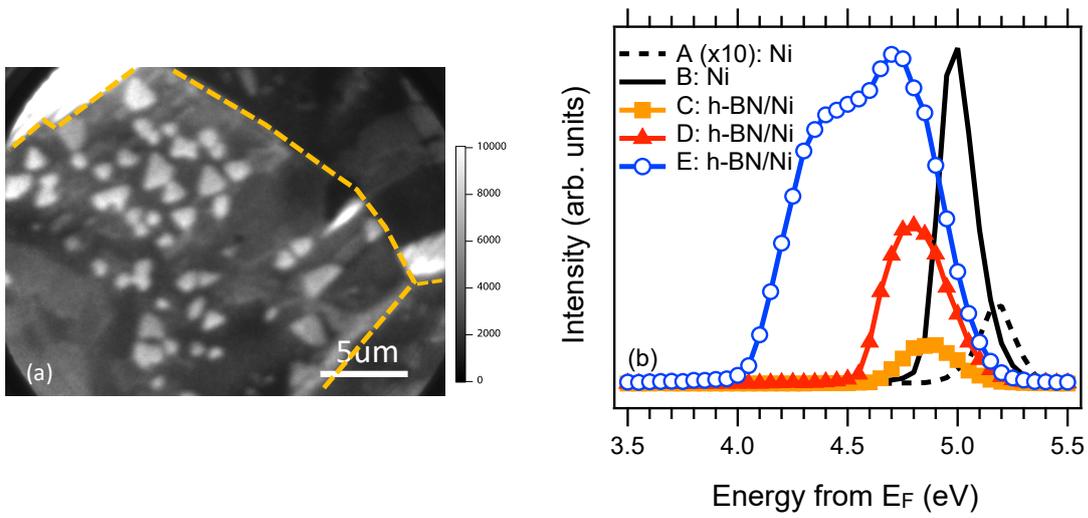

Fig. 5. (a) PEEM image of submonolayer h-BN/Ni. (b) Secondary electron spectra obtained from bare Ni surfaces (A and B) and h-BN/Ni (C, D, and E). In (a), the grain boundaries are denoted by the dashed lines.

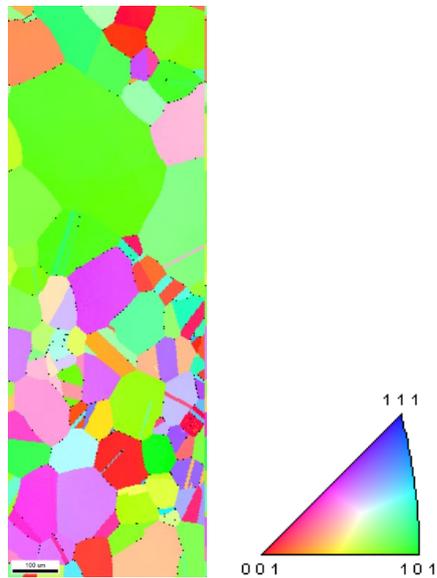

Fig. 6. EBSD mapping in *z*-direction. Scale bar: 100 μm.

Supplementary Data

## Quasi-free-standing monolayer hexagonal boron nitride on Ni

Satoru Suzuki[1], Yuichi Haruyama[1], Masahito Niibe[1], Takashi Tokushima[1], Akinobu Yamaguchi[1], Yuichi Utsumi[1], Atsushi Ito[2], Ryo Kadowaki[3], Akane Maruta[3], and Tadashi Abukawa[3]

[1]Laboratory of Advanced Science and Technology for Industry, University of Hyogo, 3-1-2 Koto, Kamigori, Ako, Hyogo 678-1205, Japan
[2]School of Engineering & Graduate School of Engineering, University of Hyogo, 2167 Shosha, Himeji, Hyogo 671-2280, Japan
[3]Institute of Multidisciplinary Research for Advanced Materials, Tohoku University, 2-1-1 Katahira, Aoba, Sendai, Miyagi 980-8577, Japan

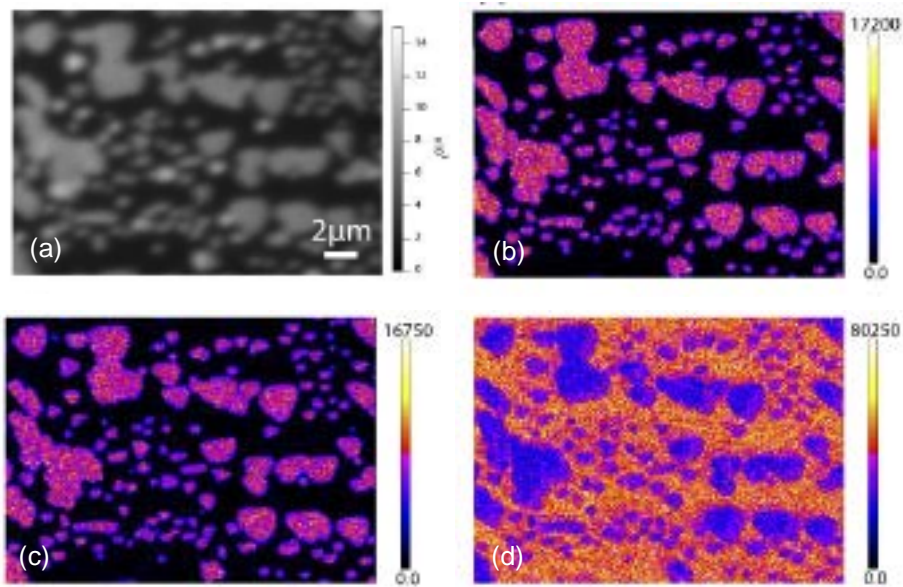

Figure S1. (a) PEEM image, (b) B-*KLL*, (c) N-*KLL*, and (d) C-*KLL* Auger mapping images of h-BN domains grown on a Ni substrate. The uniform B-*KLL* and N-*KLL* intensities mean that the h-BN domains are monolayer. In (d), the C-*KLL* signal is caused by contaminant adsorbed on the surface.